\documentstyle[preprint,tighten,aps]{revtex}

\newcommand{\resetcounter}{\setcounter{equation}{0}}     % set counter to zero
\begin{document}
%------------------------------------------------------------------------------

%-------------------------TITLEPAGE-------------------------------------

\draft
\preprint{DAMTP-98-20, hep-th/9803215}
\date{March 1998}
\title{Quantum Probes of Repulsive Singularities \\ 
        in N = 2 Supergravity}
\author{Ingo Gaida\footnote{E-mail: I.W.Gaida@damtp.cam.ac.uk, 
Research supported by Deutsche Forschungsgemeinschaft (DFG).}
}

\address{ Department of Applied Mathematics and Theoretical Physics \\
          University of Cambridge, Silver Street, Cambridge CB3 9EW, UK}
\maketitle\

\begin{abstract}
Repulsive singularities (``repulsons'')
in extended supergravity theories are investigated. 
These repulsive singularities are related to attractive
singularities (``black holes'') in moduli space of 
extended supergravity vacua.
In order to study these repulsive singularities a
scalar test-particle in the background of a repulson
is investigated. It is shown, using a partial wave
expansion, that the wave function of the
scalar particle 
vanishes at the curvature singularity at the origin. In addition the
connection to higher dimensional
$p$-brane solutions including anti-branes is discussed.
\end{abstract}

%-----------------END OF TITLEPAGE----------------------------------------

%\setcounter{page}{1}

%---------------------------------------------------------------
%
% FILE FOR PAPER               STRING BLACK HOLES
%-----------------------------------------
%
%---------------------------------------------------------------

\pacs{ \\
PACS: 04.70; 04.65; 03.67.L;
\\
Keywords: Supergravity, Physics of Black Holes, Quantum Computation.}

\newpage

%-----------------END OF TITLEPAGE----------------------------------------
%-----------------------------------------------------
%
\section{Introduction}
\resetcounter

In the last years there has been a lot of progress
in understanding black hole physics in supergravity and
string theory in $N>1$ supersymmetric vacua.
However, most of the corresponding black hole solutions
have a ``dual'' gravitational repulsive solution in moduli space.
Although the existence of these repulsive singularities is known
\cite{Linde,massless},
the repulsive singularities themselves have not been studied much, yet.
The main purpose of this article is to consider these gravitational
repulsive supersymmetric singularities\footnote{In 
\cite{Linde} these solutions were also called ``white'' holes,
but here we will call them ``repulsons'' in order to stress
that there appears an additional curvature singularity in comparison
to the ``dual'' black holes in moduli space.} (``repulsons'')
and the nature of the corresponding
antigravity effects\footnote{In general it is known for a long time that
antigravity occurs in extended supergravity 
\cite{scherk}.}. In particular a
special class of $N=2$ supersymmetric BPS repulsons
is studied and a general  
quantum mechanical analysis of 
a scalar test-particle in the background of these repulsons
is given by the use of a partial wave expansion. 
\\
The article is organized as follows: In section two static $N=2$
supergravity solutions preserving $1/2$ of $N=2$ supersymmetry are
introduced. Special points/lines in moduli space where these
solutions become either black holes or repulsons are investigated.
In section three a scalar test-particle in the background of a 
repulson is studied. In section four the connection to higher-dimensional
$p$-brane solutions is discussed and a conclusion can be found in
section five. 

%-----------------------------------------------------
%------------------------------------------------

\section{Static Solutions of N = 2 Supergravity}
\resetcounter

%------------------------------------------------
The vector couplings of local N = 2 supersymmetric Yang-Mills theory
are encoded in the holomorphic function $F(X)$ \cite{sp}, where the $X^{I}$
($I=0 \ldots N_{V}$) denote the complex scalar fields of the vector
supermultiplets. Here $N_{V}$ counts the number of physical scalars, 
and $I$ counts the number of physical vectors. 
The $N_{V}$ physical scalars
parametrize a $N_{V}$ dimensional complex hypersurface, defined by the
condition that the periods satisfy a symplectic constraint.
This hypersurface can be described in terms of a complex projective space
with coordinates $z^{A}$ ($A=1, \ldots N_{V}$), if the complex coordinates
are proportional to some holomorphic sections $X^{I}(z)$ of the 
complex projective space:
$X^{I} = e^{K(z,\bar z)/2} X^{I}(z)$
with
\begin{eqnarray}
K(z,\bar z) &=& - \mbox{log} 
\left (
i \bar X^{I}(\bar z) F_{I}( X^{I}(z) ) 
- i X^{I}(z) \bar F_{I} (\bar X^{I}(z)) 
\right ). 
\end{eqnarray}
Moreover one can introduce special coordinates
$X^{0}(z) = 1$ and $X^{A}(z)= z^{A}$
%In this special coordinates the K\"ahler potential is
%\begin{eqnarray}
%K(z,\bar z) &=& - \mbox{log} 
%\left ( 
%2 ( { \cal F} + { \bar{\cal F}} )
%- ( z^{A} - \bar z^{A} )( {\cal F}_{A} + { \bar{\cal F}}_{A})
%\right ) 
%\end{eqnarray}
with $ {\cal F}(z) = i (X^{0})^{-2} F(X)$.  
%---------------------------------------------------------------------
\\
It has been shown in \cite{FKS,FerKal1} that the 
black hole entropy \cite{Hawking} in $N=2$ supergravity 
is a topological quantity and given by 
the central charge $Z$, if the central charge has been minimized
\cite{Gibbons}
with respect to the moduli ($\partial_{A}|Z| = 0$). 
Moreover, Behrndt, L\"ust and Sabra have 
shown \cite{BLW} that one can find general stationary BPS solutions
in $N=2$ supergravity by solving the following
$2N_{V}+2$ ``stabilization equations''
\begin{eqnarray}
\label{constraint}
  X^I - \bar X^I &=& i \ H^I(r) \ = \ i 
  \left ( 
    h^I + \frac{p^I}{r}  
  \right )
\nonumber\\
F_I - \bar F_I &=& i \ H_I(r) \ = \ i 
  \left ( 
    h_I + \frac{q_I}{r}  
  \right )  
\end{eqnarray}
Here the harmonic functions $H(r)$ are given by 
the constants $h$ and the electric and magnetic charges
$q$ and $p$, respectively. 
The charges satisfy the Dirac quantisation condition
\begin{eqnarray}
\label{dirac}
  p \ q  &=& 2 \pi n, \ \ \ \ n \in {\bf Z}.
\end{eqnarray}
Moreover, we restrict ourselves here to static spherically
symmetric solutions with metric
\begin{eqnarray}
\label{metric}
  ds^2 &=& - e^{-2V(r)} \ dt^2 \ + \ e^{2V(r)} \ (dr^2 + r^2 d\Omega_2^2) 
\end{eqnarray}
Note that the metric function $e^{2V}$ is given by the holomorphic sections
\begin{eqnarray}
  e^{2V} &\equiv& i \ (\bar X^I F_I - X^I \bar F_I).
\end{eqnarray}
For black hole solutions the solution of the 
stabilization equations are on the
horizon equivalent to minimizing the central charge with respect to the
moduli. Thus, the corresponding black hole entropy \cite{Hawking} is given by
\begin{eqnarray}
\label{entropy}
  S_{BH} &=& \frac{A}{4 G_N} \ = \ \lim_{r \rightarrow 0} \ 
  \pi r^2 \  e^{2V(r)} \ = \ \pi \ |Z_{fix}|^2. 
\end{eqnarray}
In order to discuss static solutions in $N=2$ supergravity we have to
determine the prepotential. We will consider prepotentials of the form
\begin{eqnarray}
 F(X)  &=& C_{IJK} \frac{X^I X^J X^K}{X^0}
\end{eqnarray}
with intersection numbers $C_{IJK}$. In particular
we define special coordinates
$(S,T,U)=-i z^{1,2,3}$ and will consider the prepotential
\begin{eqnarray}
\label{prep}
 {\cal F}(S,T,U)  &=& -STU \ - \ a \ U^3
\end{eqnarray}
Here we keep the parameter $a$ to be an arbitrary real number. However, for
$a=1/3$ this prepotential corresponds to the heterotic
$S$-$T$-$U$ model with constant hypermultiplets
in the large moduli limit.
In this model $a$ parametrizes additional perturbative
quantum corrections of the prepotential 
(see \cite{gaida1,gaida2,rev} and reference therein).
Moreover, the microscopic interpretation, the higher order
curvature corrections and the near-extremal approximation
of this class of $N=2$ models has been studied extensively
\cite{further}.
For simplicity we take all moduli to be axion-free and
$X^0 = \bar X^0$. Solving the stabilisation equations with
respect to these constraints yields
\begin{eqnarray}
 S,T,U  &=&  \frac{H^{1,2,3}}{2 X^0}, \hspace{2cm}  
 X^0 \ = \ \frac{1}{2} \ \sqrt{-D/H_0}
\end{eqnarray}
with $D=H^1 H^2 H^3 + a (H^3)^3$.
At generic points in moduli space
this solution can represent black holes with entropy
\begin{eqnarray}
 S_{BH}  &=& 2\pi \ \sqrt{|q_0| \Delta}. 
\end{eqnarray}
Here we take $q_0<0$ and $\Delta = p^1 p^2 p^3 + a (p^3)^3$.
On the other hand it is possible to choose the parameters
of the harmonic functions in such a way that the solutions
correspond to {\em repulsive} singular supersymmmetric states
\cite{FKS}. 
In the following we will study one particular class
of these repulsons previously studied by Kallosh and Linde
in $N=4$ supersymmetric string theory \cite{Linde}.
Thus, we consider points/lines in moduli space where the metric function
\begin{eqnarray}
 e^{4V(r)}  &=& - 4 \ H_0 \ D \ = \ \sum_{n=0}^{4} \ \frac{c_n}{r^n} 
\end{eqnarray}
is a polynomial in $1/r$ to second order. Thus, the parameters of the 
harmonic functions are such that $c_{3,4}=0$. One particular example
corresponds to $a=0$ and two vanishing charges. Moreover there exist
one example with non-vanishing charges, which we will discuss in the 
following. If
\begin{eqnarray}
\label{cond1}
 H^1 H^2 \ + \ a (H^3)^2 &=& c, \ \ \ \ \ \ c=\mbox{const}
\end{eqnarray}
i.e. one harmonic function is a function of the other two, the condition
$c_{3,4}=0$ is also satisfied. In string theory this might correspond to a 
gauge symmetry enhancement point/line in moduli space 
where additional massless states occur.
If we take $h=1$ and $H_0 = - (1 + \frac{q_0}{r}) $
the condition (\ref{cond1}) reads in terms of the charges
\begin{eqnarray}
\label{cond2}
 p^1 p^2 \ + \ a (p^3)^2 = 0,
\hspace{1cm}
 p^1 + p^2 + 2a p^3 = 0
\end{eqnarray}
These conditions can be solved and yield
\begin{eqnarray}
 p^1_{\pm} &=& - (1+2a) p^2 \ \pm \ 2 p^2 \sqrt{a+a^2}.
\end{eqnarray}
Since all charges are real this yields the bound
$a^2 + a \geq 0$.
If the charges satisfy these equations it follows $c=1+a$. Hence,
after a suitable coordinate transformation the metric function becomes
\begin{eqnarray}
 e^{4V(r)}  &=& 1 + \frac{q_0 + p^3}{r} +  \frac{q_0 p^3}{r^2} 
\end{eqnarray}
The ADM mass of this solution is $M=(q_0 + p^3)/4 \geq 0$.
From the Dirac quantisation condition (\ref{dirac}) follows
$q_0p^3 = 2\pi n$ with integer $n$. The solution 
corresponds to a black hole if
$n \geq 0$ and to a repulson if $n< 0$. Moreover,
for $q_0= -p^3$ this solution represents a massless 
repulson. Note that the entropy of these solutions vanishes, although
all harmonic functions and charges are non-vanishing.

%------------------------------------------------

\section{Scalar Particles in the Background of a Repulson}
\resetcounter

%------------------------------------------------

Following \cite{Linde} 
we will consider the motion of a scalar test-particle in the
background of a repulson, i.e. the metric reads
\begin{eqnarray}
 e^{4V(r)}  &=& 1 + \frac{4M}{r} - \frac{4 Z^2}{r^2},
\hspace{1cm} M,Z \geq 0.  
\end{eqnarray}
The corresponding Ricci tensor is given by
\begin{eqnarray}
R_{tt}  &=& -6 \ 
            \frac{Mr^3 + 3 M^2 r - 3 r^2 Z^2 - 12 M r Z^2 + 8 Z^2}
                 {(r^2+4Mr-4Z^2)^2 \ r^2}
\nonumber\\
R_{rr}  &=& R_{\phi \phi} \ = \  R_{\theta \theta}
\ = \ \frac{6Z^2-2Mr}{r^4}.
\end{eqnarray}
Thus, the metric has one curvature singularity at $r=0$ and another
``naked'' curvature singularity at 
\begin{eqnarray}
 r_0  &=& 2 
 \left ( 
  \sqrt{M^2 + Z^2} - M 
 \right )
\end{eqnarray}

%-------------------------------------------
\subsection{Classical Analysis}
%--------------------------------------------
In the classical limit the Newtonian potential $\Phi$ is given by
\begin{eqnarray}
 \Phi(r)  &=& - \frac{1}{2} \ (g_{tt} + 1) \ = \  
               - \frac{M}{r} + \frac{Z^2}{r^2}.
\end{eqnarray}
The corresponding strength of the gravitational field
$\Phi^\prime = \frac{M}{r^2} - \frac{2Z^2}{r^3}$ is gravitational
attractive at large distances ($r>r_c$) and gravitational repulsive
for $r<r_c$. The critical distance where gravitational repulsion and
attraction yield a vanishing net force is given by $r_c = 2 Z^2/M$. For
massless repulsons the Newtonian potential is always repulsive.
Using Hamilton-Jacobi theory (see \cite{Landau1} for example) 
one can show that a test particle of small mass $m$ ($M >> m$), energy $E$ at
$r \rightarrow \infty$ and angular momentum $L$
needs the following time to move from $r_1$ to $r_2$ 
\begin{eqnarray}
 t  &=& \int_{r_1}^{r_2} \ dr \ 
 \frac{e^{4V}}{\sqrt{E^2 e^{4V} - \frac{L^2}{r^2} - m^2 e^{2V}}}
\end{eqnarray}
It follows that for $L=0$ a massive test-particle becomes reflected
by the repulson at 
\begin{eqnarray}
 r_{min}  &=& \frac{2}{\epsilon}  
 \left ( 
  \sqrt{M^2 + \epsilon Z^2} - M 
 \right ) \ > \ r_0,
 \hspace{1cm} \epsilon = 1 - \frac{m^4}{E^4} 
\end{eqnarray}
Moreover, for $L \neq 0$ a massless test-particle becomes reflected
by the repulson at 
\begin{eqnarray}
 r_{min}  &=& 2
 \left ( 
  \sqrt{M^2 + Z^2 + \frac{L^2}{4E^2} } - M 
 \right ) \ > \ r_0.
\end{eqnarray}
Note that for a massless repulson this classical analysis is not valid,
since we have chosen the mass of the repulson large, so that the
center of mass of the two-body problem is given by the repulson. 
Moreover, for a massless test-particle in the s-state ($m=L=0$) a
quantum mechanical analysis analogous to \cite{Holzhey,Marolf} is necessary.

%-------------------------------------------
\subsection{Quantum Mechanical Analysis}
%--------------------------------------------

We consider a scalar test-particle $\psi$ of mass
$m$ satisfying the Klein-Gordon equation in the background
of the repulson:
\begin{eqnarray}
\label{KG}
 \partial_\mu \
 \left (
  \sqrt{-g} \ g^{\mu \nu} \ \partial_\nu \ \psi
 \right )
   &=& - m^2 \ \psi
\end{eqnarray}
Expanding $\psi$ in partial waves 
$\psi= e^{-i \omega t} R_{kl}(r) Y_{lm}(\theta,\phi)$
(\ref{KG}) becomes the Schr\"odinger equation 
\begin{eqnarray}
\label{SCH}
 \Delta_r \psi  \ + \ (E - V) \psi &=& 0
\end{eqnarray}
with
\begin{eqnarray}
 E &=& m^2 + \omega^2,
\nonumber\\
 \Delta_r &=& \frac{\partial^2}{\partial r^2} + 
            \frac{2}{r} \frac{\partial}{\partial r},
\nonumber\\
 V (r) &=& - \frac{4M\omega^2}{r} + \frac{4Z^2\omega^2 + l(l+1)}{r^2}.         
\end{eqnarray}
The Schr\"odinger equation (\ref{SCH}) can be solved in the 
quasi-classical regime, near the singularities $r=0$ and $r=r_0$ and
throughout the entire space-time using standard techniques 
(see \cite{Landau2} for example). First we will consider (\ref{SCH})
in the quasi-classical regime for small $r$: 
The de Broglie wavelength is given by $\lambda = \frac{r}{s(s+1)}$ 
with\footnote{Note that $s$ is quantized in $n$ and $l$ 
for the solution discussed in section II.}
$s(s+1)=4Z^2\omega^2 + l(l+1)$. The quasi-classical regime
requires $| \frac{d \lambda}{dr}| = \frac{1}{\sqrt{s(s+1)}} << 1$.
Thus, for large $s$, i.e. large angular momentum of $\psi$ or
strong gravitational repulsion with $Z^2 << M$, 
the quasi-classical approximation is
valid. In this approximation the radial part of the wave function becomes
\begin{eqnarray}
 \psi &\sim& r^s         
\end{eqnarray} 
and has no singular behaviour as one approaches the singularity
at $r=r_0$. Thus, the singularity at $r=r_0$ is transparent for
massive or massless scalar particles at the quantum level.
This result is quite general and does not rely on the partial wave
expansion of the scalar test-particle\footnote{I thank G. Gibbons
for pointing this out to me.}, since the Klein-Gordon equation
of the scalar test-particle
is in general non-singular at $r=r_0$ for the repulson background.  
\\
Following \cite{Linde} we will show now that
the scalar test-particles are totally reflected at the
singularity at $r=0$. In order to study this effect
the ingoing wave function of the scalar must be able to
tunnel through the singularity at $r=r_0$.
Here we ``suggest''
the following generalization of the metric (\ref{metric}) for
$r>0$ 
\begin{eqnarray}
\label{gen_metric}
  ds^2 &=& - e^{-4V(r)} \ |e^{2V(r)}|
           \ dt^2 \ + \ |e^{2V(r)}| \ (dr^2 + r^2 d\Omega_2^2).
\end{eqnarray}
Thus, at $r=r_0$ the determinant of the metric changes its sign.
The corresponding ``suggested'' metric for $0<r<r_0$ reads
\begin{eqnarray}
\label{nmetric}
  ds^2_{r<r_0} &=& 
  |e^{-2V(r)}| \ dt^2 \ + \ |e^{2V(r)}| \ (dr^2 + r^2 d\Omega_2^2) 
\end{eqnarray}
For small $r<r_0$ one can take $E-V \sim \frac{s(s+1)}{r^2}$.
Moreover, from the quasi-classical analysis follows that
the radial part of the wave function $\psi$ can be approximated
by a polynomial in $r$, i.e. $R_{kl} = A r^n$. Solving
the Schr\"odinger equation (\ref{SCH}) in this regime
yields 
\begin{eqnarray}
  n_\pm &=& \frac{1}{2}  
  \left (
  \pm \sqrt{1 + 4s(s+1)} -1
 \right )  
\end{eqnarray}
From the quasi-classical analysis follows already that $n_+$
is the correct choice for the parameter $n$. However, we will proof this
now in general: One can consider a small region of radius $\rho$
around the origin and replace $s(s+1)/r^2$ by $s(s+1)/\rho^2$
in the Schr\"odinger equation (\ref{SCH}) for small $r$. 
Solving now 
\begin{eqnarray}
 \Delta_r \tilde R  \ - \ \frac{s(s+1)}{\rho^2} \tilde R   &=& 0
\end{eqnarray}
yields
\begin{eqnarray}
 \tilde R(r) &=& C \frac{\sin (i \tilde k r)}{r},
 \hspace{1cm}
 \tilde k^2 = s(s+1)/\rho^2.
\end{eqnarray}
Note that $\tilde R$ is finite and vanishing at $r=0$.
Now we take as general ansatz $R(r)=A r^{n_+} + B r^{n_-}$.
Since $R$ and $\tilde R$ and their derivatives must be continuous,
it follows
\begin{eqnarray}
 \partial_r \ \log (r \tilde R)_{|r=\rho} &=& 
 \partial_r \ \log (r R)_{|r=\rho}
\end{eqnarray}
This condition yields
\begin{eqnarray}
 \frac{B}{A} &=& - \ \frac{i \sqrt{s(s+1)} \cot(i\sqrt{s(s+1)}) - n_+ -1}
                          {i \sqrt{s(s+1)} \cot(i\sqrt{s(s+1)}) - n_- -1}
                          \ \rho^{n_+-n_-}.
\end{eqnarray}
It follows 
$ \lim_{\rho \rightarrow 0} B/A = 0$.
Thus, from
\begin{eqnarray}
 \lim_{\rho \rightarrow 0} \ \frac{1}{A} \ R(\rho) &=& 
 \lim_{\rho \rightarrow 0} \ \frac{1}{A} \ \tilde R(\rho) \ = \ 0
\end{eqnarray}
follows for the radial part of the scalar wave function for small $r$:
$R(r) = C r^{n_+}$ with some constant $C$. Hence, the scalar wave function
vanishes at the singularity $r=0$. This result is valid for massive and 
massless repulsons.
\\
To investigate the space-time geometry near the reflecting singularity
we consider the metric (\ref{nmetric}) and perform a Weyl-transformation
of the metric $g_{\mu\nu} \rightarrow |e^{2V}| g_{\mu\nu}$ and an
additional coordinate transformation $r^\prime = - 2 Z \log r$. This
yields the following metric near the singularity $r=0$.
\begin{eqnarray}
  ds^2 &=& dt^2 \ + \ dr^{\prime 2} + 4Z^2 d\Omega_2^2. 
\end{eqnarray} 
Thus, the four-dimensional space-time ${\cal M}_4$ near the repelling
singularity is in this particular ``frame'' a product space
$ {\cal M}_4  = {\bf R^2} \times {\bf S^2}$,
where the radius of  ${\bf S^2}$ is associated to the strength of the
gravitational repulsion. It follows for the solution discussed in
section II that the radius of ${\bf S^2}$
and the corresponding ``area'' $A = 16 \pi Z^2$ are quantized, since
$4 Z^2 = 2 \pi |n|$.
\\
Now we will solve the Schr\"odinger equation (\ref{SCH}) in the region
$r>r_0$. Introducing the wave number $k=\sqrt{E}$ and
$\alpha=2M\omega^2$ the radial part
of the wave function is given by \cite{Landau2}
\begin{eqnarray}
 R_{kl}(r) &=& \frac{C_k}{(2s+1)!} (2kr)^s e^{-ikr}
 F(\frac{i}{k}+s+1,2s+2, \alpha ikr )
\end{eqnarray}
Here $F(a,b,x)$ denotes the confluent hypergeometric 
functions\footnote{$F(a,b,x)=
\sum_{n=0}^{\infty} \frac{a_n}{b_n} \frac{x^n}{n!}$
with $a_0=1, a_n=a(a+1) \ldots (a+n-1),
b_n = b (b+1) \ldots (b+n-1)$.
} and the constant $C_k$ is defined as follows
\begin{eqnarray}
 C_k &=& \sqrt{\frac{2}{\pi}} \ k \ e^{\frac{\alpha \pi}{4k}}
         |\Gamma(s+1- i \frac{\alpha}{2k})|,
\hspace{1cm}
\int dr \ r^2 R_{kl} R_{k^\prime l} = 2 \pi \delta ({k-k^\prime}).
\end{eqnarray}
For $r \rightarrow \infty$ this solution becomes
\begin{eqnarray}
 R_{kl}(r) &=& \sqrt{\frac{2}{\pi}} \ \frac{1}{r} \
 \sin (kr + \frac{\alpha}{2k} \log (2kr) - \frac{s \pi}{2} + \delta_s),
\hspace{1cm}
\delta_s = \mbox{arg} \ \Gamma(s+1-i \frac{\alpha}{2k}). 
\end{eqnarray}
This result implies that the scattering process 
of a scalar particle of mass $m$ and a heavy repulson of mass
$M >> m$ is elastic, only, and the inelastic
cross-section vanishes. Hence, in a scattering process
the repulson reflects any scalar test-particle. Note that this conclusion
holds for a massive and massless scalar test-particle, but only
for a massive and heavy repulson.  

%------------------------------------------------

\section{Repulsons from Extremal P-Branes}
\resetcounter

%------------------------------------------------

The class of repulson solutions we have studied so far can be
associated to higher dimensional brane configurations. In 
general they correspond to orthogonally intersecting 
brane-anti-brane configurations. These configurations become
massless if the absolute values of the 
corresponding two brane-charges are equal. 
If we consider, for example, four-dimensional configurations 
in $N \geq 2$ supergravity with a prepotential of the
form (\ref{prep}) with $a=0$ and
non-vanishing charges, then the particular solution can be
interpreted as the intersection of three 4-branes and one
0-brane in $D=10$ type IIA string theory. In M-theory
this solution correspond to the intersection of three 5-branes
with a boost along the common string \cite{Tseytlin_1,BLW}.
Wrapping the 5-branes around 4-cycles this configuration 
corresponds to a magnetic string in five dimensions
(see e.g. \cite{Behrndt,gaida2} and reference therein).
Additional wrapping around the 5th direction yields the
four dimensional solution. The appearance of negative charges
defines the corresponding anti-brane configurations and yields
massless solutions at particular points in moduli/charge space
\cite{massless,BLW,gaida2}. Note that solutions with anti-brane
intersections, only, have 
negative ADM mass (see also \cite{negative}).   
Switching off two charges corresponds to two vanishing branes and
studying the corresponding brane-anti-brane intersection in four
dimensions yields the repulson background studied above \cite{Larsen}.  

%-----------------------------------------------------------------
\section{Summary, Discussion and Conclusion}
\resetcounter

Supersymmetric BPS solutions of $N=2$ supergravity have been studied.
In particular black holes and repulsons have been considered and it has
been shown that both types of singularities are closely related to
each other, i.e. a black holes solution at one point in moduli space
becomes a repulson at another point
\cite{massless}. For a two-charge configuration, for 
example,
with a generic choice for the moduli at infinity and positive greater 
charge, the
lower charge with a plus sign corresponds to a black hole solution and
the lower charge with a minus sign to a repulson. It might be interesting
to investigate whether there are as many repulsive as attractive solutions in
moduli space, i.e. almost all attractive black hole solutions might have  
``dual'' repulsive solutions. 
Since a repulson is gravitational attractive at long distances and 
gravitational repulsive at short distances, it is particle-like
\cite{Weyl,Holzhey,Larsen}. In order to investigate the reflecting
nature of a repulson a scalar test-particle in the 
background of a repulson has been studied \cite{Linde}. 
It has been shown in general
that the wave function of the scalar vanishes at the singularity
$r=0$ and is transparent at the singularity at $r=r_0$, if the wave
function can be expanded in partial waves.
Moreover, any scattering process of a heavy repulson and a scalar
is elastic, only. These results suffer from the
boundary condition at the naked singularity at $r=r_0$. Here we
``suggested'' a generalization of the canonical metric valid for
$r>0$ to study tunneling of the ingoing scalar test-particle through
the naked singularity at $r=r_0$. It would be very interesting
to investigate this point further. Moreover,   
massless repulsons \cite{massless,gaida2}
are rather special and also deserve further investigations. 
Note that the second part of this article can be applied to the
particular string model with non-Abelian gauge fields
studied by Kallosh and Linde
in \cite{Linde} and generalizes their
quantum mechanical analysis of a scalar test-particle in the
background of a repulson.
\\
From the brane point of view the class of repulsons
studied here correspond to orthogonally brane-anti-brane intersections. The
corresponding ``dual'' black hole solution in moduli space would correspond
to the brane-brane intersection. Moreover, the 
corresponding anti-brane-anti-brane intersection is also gravitational
attractive, but its ADM mass is negative. In general the four
dimensional solutions are
gravitational repulsive if they contain an odd number of anti-branes and 
gravitational attractive if the number of anti-branes is even. 
\\
To conclude, naked repulsive singularities appearing in extended 
supergravity models have been studied in great detail.
These repulsive singularities are related to black hole solutions 
in moduli space \cite{massless},
i.e. in moduli space 
these repulsive singular solutions are as generic as their attractive
``duals''. 
It would be very interesting to investigate more
general repulsive singular solutions
with, for instance, additional $1/r^3$ and $1/r^4$ 
contributions to the metric function or without using a partial wave
expansion.  

%------------------------------------------------------------------
\bigskip  \bigskip

\noindent
{\bf Acknowledgments}  \medskip \newline
I would like to thank T. Mohaupt, K. Behrndt, T. Turgut, H. Hollmann, 
G. Papadopoulos and G. Gibbons for
discussions and F. Larsen for correspondence.

%--------------------------------------------------------------------

%---------------------------------------------------------------------
%
%                        BIBLIOGRAPHY
%
%----------------------------------------------------------------------

\renewcommand{\arraystretch}{1}

\newcommand{\NP}[3]{{ Nucl. Phys.} {\bf #1} {(19#2)} {#3}}
\newcommand{\PL}[3]{{ Phys. Lett.} {\bf #1} {(19#2)} {#3}}
\newcommand{\PRL}[3]{{ Phys. Rev. Lett.} {\bf #1} {(19#2)} {#3}}
\newcommand{\PR}[3]{{ Phys. Rev.} {\bf #1} {(19#2)} {#3}}
\newcommand{\IJ}[3]{{ Int. Jour. Mod. Phys.} {\bf #1} {(19#2)}
  {#3}}
\newcommand{\CMP}[3]{{ Comm. Math. Phys.} {\bf #1} {(19#2)} {#3}}
\newcommand{\PRp} [3]{{ Phys. Rep.} {\bf #1} {(19#2)} {#3}}

%-----------------------------------------------------------------------------

\begin{thebibliography}{9}


\bibitem{Linde} R. Kallosh and A. Linde, \PR{D 52}{95}{7137}

\bibitem{massless}
K. Behrndt, \NP{B455}{95}{188};
\\
M. Cveti{\v c} and D. Youm, Phys. Lett. {\bf B359} (1995) 87;
\\
C.M. Hull, Proceedings of Strings `95, World Scientific,
1996;
\\
K.-L. Chan and M. Cveti{\v c}, Phys. Lett. {\bf B375} (1996) 98;
\\
T. Ortin, Phys. Rev. Lett. {\bf 76} (1996) 3890; 
\\
M. Cveti{\v c} and A.A. Tseytlin, Phys. Lett. {\bf B366} (1996) 95;
\\
K. Behrndt, D. L\"ust and W.A. Sabra, {\tt hep-th/9708065};
\\
M. Cveti{\v c} and C.M. Hull, {\tt hep-th/9709033}; 

\bibitem{scherk} 
J. Scherk, Phys. Lett. {\bf B88 }
(1979) 265. 
\\
S. Bellucci, ``1260: Phenomenology of Antigravity in N=2,8
Supergravity'', Talk presented at the International Conference
on High Energy Physics, Jerusalem, Israel, 19-26. Aug. 1997,
and reference therein.

\bibitem{sp}
B. de Wit, P.G. Lauwers and A. Van Proeyen, Nucl. Phys. {\bf B 255 }
(1985) 569.
\\
B. de Wit, F. Vanderseypen and A. Van Proeyen, Nucl. Phys. {\bf B400}
 (1993) 463.


\bibitem{BLW} K. Behrndt, D. L\"ust and W.A. Sabra,
\NP{B510}{98}{264}.
%{\tt hep-th/9705169}.
 
\bibitem{Hawking}
J. Bekenstein,
\PR{D7}{73}{2333};
\\
S. Hawking, Comm. Math. Phys. {\bf 43} (1975) 199.
 
\bibitem{FKS}
S. Ferrara, R. Kallosh and A. Strominger,\PR{D 52}{95}{5412} 

\bibitem{gaida1}
I. Gaida, {\tt hep-th/9705150}, to appear in Nucl. Phys. B.

\bibitem{gaida2}
I. Gaida, {\tt hep-th/9802140}

\bibitem{rev}
D. L\"ust, {\tt hep-th/9803072}
 
\bibitem{FerKal1} S. Ferrara and R. Kallosh,
\PR{D 54}{96}{1514},\PR{D 54}{96}{1525}

\bibitem{Landau1} L.D. Landau and E.M. Lifshits,
The Classical Theory of Fields, Pergamon Press (1975).

\bibitem{Holzhey} C. Holzhey and F. Wilczek, \NP{B380}{92}{447}

\bibitem{Marolf} G.T. Horowitz and D. Marolf,
\PR{D 52}{95}{5670}
% {\tt gr-qc/9504028}

\bibitem{Landau2} L.D. Landau and E.M. Lifshits,
Quantum Mechanics, Pergamon Press (1977).

\bibitem{Tseytlin_1} 
A.A. Tseytlin, \NP{B475}{96}{149};
\\
G. Papadopoulos and P.K. Townsend, 
Phys. Lett. {\bf B380} (1996) 273. 

\bibitem{Behrndt}
K. Behrndt, Phys. Lett. {\bf B396} (1996) 77,
\\
K. Behrndt and T. Mohaupt, \PR{D56}{97}{2206}.

\bibitem{Larsen}
V. Balasubramanian and F. Larsen, \NP{B495}{97}{206}; 


\bibitem{further}
J. Maldacena, A. Strominger and E. Witten, JHEP  {\bf 12} (1997) 002;
% {\tt hep-th/9711053}; 
\\
K. Behrndt, M. Cveti{\v c} and W.A. Sabra, {\tt hep-th/9712221};
\\
K. Behrndt, G.L. Cardoso, B. de Wit, D. L\"ust, T. Mohaupt
and W.A. Sabra, {\tt hep-th/9801081};

\bibitem{Weyl}
H. Reissner, Ann. Phys. 50 (1916) 106;
\\
H. Weyl, Sitz. Ber. Preuss. Akad. Wiss. 465 (1918)
\\
F. Finster, J. Smoller and S.T. Yau, {\tt gr-qc/9802012}; ;


\bibitem{Gibbons} S. Ferrara, G. W. Gibbons and R. Kallosh, 
\NP{B500}{97}{75},
%{\tt hep-th/9702103}.

\bibitem{negative} 
M. Mann, Class. Quant. Grav.  {\bf 14} (1997) 2927; 
\\
M.S. Costa and G. Papadopoulos,
\NP{B510}{98}{217};

\end{thebibliography}
\end{document}